# Accountability in Open Source Software Ecosystems: Workshop Report

**October 2025, Carnegie Mellon University**

Nandini Sharma, Thomas Bock, Rich Bowen, Sayeed Choudhury, Brian Fitzgerald, Matt Germonprez, Jim Herbsleb, James Howison, Tom Hughes, Min Kyung Lee, Stephanie Lieggi, Andreas Liesenfeld, Georg Link, Nicholas Matsakis, Audris Mockus, Narayan Ramasubbu, Christopher Robinson, Gregorio Robles, Nithya Ruff, Sonali Shah, Igor Steinmacher, Bogdan Vasilescu, Stephen Walli, Christopher Yoo.

**\*Author Affiliations available in the Appendix**

# Table of Contents

# Executive Summary

Open source software ecosystems are composed of a variety of stakeholders including but not limited to non-profit organizations, volunteer contributors, users, and corporations. The needs and motivations of these stakeholders are often diverse, unknown, and sometimes even conflicting given the engagement and investment of both volunteers and corporate actors. Given this, it is not clear how open source communities identify and engage with their stakeholders, understand their needs, and hold themselves accountable to those needs. We convened 24 expert scholars and practitioners studying and working with open source software communities for an exploratory workshop discussion on these ideas. The workshop titled "Accountability and Open Source Software Ecosystems" was organized on Oct 14–15 on campus in Carnegie Mellon University, Pittsburgh, PA. The purpose of this in-person workshop was to initiate conversations that explore important and urgent questions related to the role of accountability in open source software ecosystems, and to inspire an exciting research agenda and meaningful stakeholder engagement ideas for practitioners.

The workshop began with a keynote by Stephen Walli, Principal Program Manager at Microsoft in the Azure Office of the CTO, highlighting the role of OSS communities in the software industry, the importance of non-profits, an overview of OSS stakeholders, and an introduction of sustainability-related challenges in open source. The keynote was followed by three theme-based panels and hour-long brainstorming sessions corresponding to each panel discussion. The panels focused on the following questions respectively: **(1) Who are the stakeholders of open source software communities and how can we identify their needs? (2) How and to what extent can we rally stakeholders to consider broader concerns and which stakeholders are able to respond to and fulfill such responsibilities? (3) How can the stakeholders of open source ecosystems engage with each other? What are the ways in which meaningful interaction can be enabled with communities? And, what is the role of Open Source Program Offices (OSPOs) in doing the same?** After each panel discussion, participants formed breakout discussion groups. The discussion groups were formed based on a set of 3-5 questions raised by the workshop participants after the panel discussions. This report summarizes the ideas, questions, and recommendations that came up during the panel discussions as well as the brainstorming conversations. More details about the panel programming, the brainstorming sessions, and the workshop participants can be found in the Appendix.

# Key Themes

The workshop led to generative discussions around the following key themes. While these themes are not an exhaustive representation of the workshop discussions, they represent ideas that led to some of the most engaging conversations during the course of the workshop. For further context, please refer to the detailed description of each panel and brainstorming session.

- Engagement of corporate actors with open source communities varies significantly based on the kinds of open source communities they engage with. However, an important part of this engagement is to (a) understand how to fund communities; and (b) to be engaged, to be present, and to offer support in a variety of ways.
- Open source communities need formal front doors for outside entities (e.g., corporations, volunteers, and OSPOs) to engage with them.
- The velocity at which open source communities grow and the extent of their coverage through foundations is a signal of their health and sustainability.
- Plurality of sub-communities in a larger community can lead to conflict but the key is to find ways to communicate in a manner that is meaningful to those sub-communities.
- Some kind of accountability is necessary in open source to ensure that the product they produce is consumable.
- It is crucial to understand and broaden the meaning of a maintainer to include a diverse set of skills as representative of and required for this responsibility, and to identify that ownership of a project essentially means being in a strong relationship with the community.
- A non-profit foundation-like entity associated with relevant open-source communities is helpful in organizing funding mechanisms for community needs.
- The support and visibility of OSS supply chains is a multifaceted issue including a differentiation of visibility and transparency, consideration of sociotechnical relationships embedded in supply chains, acknowledgement of legal responsibilities of corporations, and sharing of growth-related challenges of individual OSS projects.
- Sustainability of a community should be important for a corporation intending to rally community stakeholders towards shared goals.
- OSS value does not necessarily manifest through quarterly profit but through longer-term engagement.
- When a sufficient number of community members contribute meaningfully and significantly but in unplanned ways, it indicates that collective action does not necessarily need collective direction.
- The work of corporate OSPOs is that of engaging in diplomacy and advocating for open source project support inside a company but also advancing company interests in open source communities.
- As a village becomes a town and a town becomes a city, the internal culture of implicit values and norms gives way to explicit rules and regulations, and transferring leadership becomes challenging.

# Detailed Description

## Panel #1

**Theme: Who are the stakeholders of open source software communities and how can we identify their needs?**

The panel began with a set of prompts for each panelist. After each panelist addressed the relevant prompt, the rest of the workshop participants had an opportunity to respond to that prompt or engage with the responses from the panelist for the prompt. The list of prompts included:

- If I am shaping open source policy for a commercial firm, how should I expect an open source community to engage with us?
- How can I advocate for the community to prioritize features and bug fixes I need? What are my leverage points?
- How can I tell if a project is a good candidate to include in my software supply chain?
- What level of conflict is acceptable in community interactions? Are there warning signals to steer clear?
- Are software communities accountable? Who are they accountable to, and what are they accountable for?

Discussion around the prompts mentioned above led to the following discussion items around **engagement of corporate actors with open source communities** and **identification of stakeholders**:

**Engagement of corporate actors with open source communities varies significantly based on the kinds of open source communities they engage with.** Given potentially opposing ideologies, there is a possibility of tension and conflict between corporations and communities. For example, corporations tend to have specific expectations of communities in terms of responsible behavior which may or may not be acceptable to communities. While different communities operate differently and it is hard to generalize the do's and don'ts of such an engagement, the existing landscape of open source communities suggests that many communities are benevolent dictatorships with two-level governance where the top drives community direction but also needs to incentivize the community to some extent. One of the ways in which corporations can consider engaging with communities is to not have any expectations from the community but engage with the idea of partnering and collaborating with them rather than thinking about them as a service provider. When corporations do not understand the needs of the communities and expect them to deliver features and products, corresponding associations often do not yield good results for corporations. The ability of **corporations to be able to support their own needs while staying engaged with the community** determines the success of their engagement with open source communities. Treating communities as service providers and not understanding the culture and goals of the communities deeply is a frequent mistake on part of the corporations. An important part of this

engagement is to (a) **understand how to fund communities** which could differ significantly based on the community involved and the context of corporate involvement; and (b) **to be engaged, to be present, and to offer support** in a variety of ways.

On the other hand, the **communities need formal front doors** for outside groups to engage with them. While individuals might find it easier to engage with open source, groups (e.g., corporations, non-profits, other OSS projects) often have a hard time finding relevant entry points. There are examples of communities (e.g., Open Stack) that have deliberately structured themselves in a manner that allows new members to ramp up and easily understand how they might be able to succeed. However, there can be pitfalls of establishing such structures that might seem helpful at the outset (e.g., creating a core representative group). Formal structures can create unnecessary division and conflict among members as power gets concentrated in unequal ways and with no acceptable justification. One of the ways of countering such tensions is to include more actors in the community which can help dissolve monopolies and concentration of power. For example, when it becomes difficult to deal with actors from one specific corporation, including more corporations can help. Rust has tried to address this issue by creating sub-teams that own bits of Rust functionality. The project goals program that Rust introduced has attempted to create an entry point to the community for outsiders while avoiding a structure that might distribute power in seemingly unequal ways.

In deciding **which communities corporate actors must or must not engage with** in terms of including them in their software supply chains, metrics such as the **CHAOSS metrics**[1] can be helpful. CHAOSS metrics help understand organizational corporate diversity, the direction of the community, and the ability of various actors to re-license corresponding products. However, in understanding open source communities, it is also crucial to focus on the aspects that cannot be understood well through collected or reported metrics. For example, metrics related to code contributions do not take into consideration code reviewing efforts. **Velocity at which open source communities grow** is another critical consideration in understanding whether or not the community is able to sustain itself or is dying and what is the stage of its lifecycle at any point in time. Moreover, while metrics such as those reported by CHAOSS are important indicators and determinants of the suitability of a community for software supply chains, these metrics might not be available for less mature communities because the amount of trace data available for smaller communities especially on online platforms such as GitHub is likely to be minimal. In those cases, the **coverage of a community through non-profits** such as foundations can signal community viability.

Other challenges to corporate engagement with open source include the need to **navigate the plurality of sub-communities in a larger community**. Those involved with open source can often have multiple but strong identities of being associated with both the corporate culture and the open source culture. **Being aware of the dynamics that are acceptable in different spaces** (corporate versus community driven) is crucial to navigating such an environment as this setting suggests possible conflicts in open source ecosystems. Moreover, having people who can help temper conflicting dynamics can be incredibly helpful as well.  However, when conflicts in such heterogeneous systems can not be managed, repercussions include **hard forking** of open source projects. While this may create multiple options for those who want to include open source in software supply chains, it can also be costly as it can fragment a community especially when promises for openness and transparency are not kept. Technical conflicts rarely lead to such an outcome as a hard fork, but it depends on the severity of the transgression. The ability of corporations and open source communities to show trust, empathy, and support for each other can be helpful tools in resolving such conflicts. Moreover, there is evidence to suggest that **communities are often stronger together**. When projects have forked in the past owing to pressures of business models, they have often reunited after years (e.g., GCC and EGCS, node.js and jo.js etc.). Regardless, the capacity to make important community-related decisions is dependent on Intellectual Property ownership which, if taken badly, can cause significant community crises such as these.

---

[1] "CHAOSS is an acronym for Community Health Analytics in Open Source Software and is focused on creating metrics, metrics models, and software to better understand open source community health on a global scale." (CHAOSS, n.d., https://chaoss.community/)

As far as **identification of stakeholders** is concerned and the ways they might think about accountability, there are some arguable but helpful entry points to the discussion. While, in a narrow sense, **contributors to open source communities can be considered the only stakeholders** because they are doing the work to support a community, there might be reasons to argue otherwise. Those who have reasons to hold some kind of expectations from open source communities could be argued to have stakes in their work. For example, even though the disclaimers included in many open source projects void any accountability to any stakeholder whatsoever, there is a **baseline expectation from a community or weak accountability** of sorts towards those who might depend on that product and therefore expect that it be consumable at least to a certain degree. These individuals and groups are the stakeholders of open source communities with some fundamental expectations. For example, if a software product comes with a default password, the baseline expectation is that the password must be customizable. While this is a simplistic example, other stakeholder expectations could include **transparency and visibility into product development** processes as they can help provide some form of consumer protection. Among those who need to actively consider implications of working with open source from a liability perspective are foundations and non-profit organizations who own trademarks of open source communities and could be considered its stakeholders. Their expectations and contributions to open source could be vastly different. For example, the Linux foundation cares about the success of the Rust For Linux project for a variety of reasons and has therefore needed to develop a long term relationship with the project so as to be able to meaningfully support it.

## Brainstorming Session #1

The panel #1 discussions mentioned above generated following topics for breakout discussion sessions among the participants. Please refer to the Appendix for further details on questions discussed during the brainstorming session.

**Open source maintenance as a career: How to make it happen? How to generate sustainable funding? What is a reasonable expectation of a maintainer of an OSS project?**

Thinking about open source maintenance as a career raises the fundamental **question regarding the definition and role of a maintainer**. For example, a maintainer can be understood as someone who has created a project and is responsible for the success of that project where the term "success" can be defined broadly especially given the variety of ways maintenance work continues to be done. New members to a community can also be considered maintainers as they would eventually graduate, over a period of time, to positions of responsibility where they would have tasks clearly demarcated as maintenance-related. Open source communities have demonstrated maintainer turnover where a maintainer who starts a project explicitly passes it on to those who came after. As far as specific tasks are concerned, funding for maintenance-related work, especially through foundations, points to the importance of both community engagement and product maintenance work as both are funded to a certain degree by community foundations. Both community maintenance and product maintenance are equally necessary and important parts of maintainers' work. It also indicates that any kind of ownership that a maintainer or a group of maintainers can claim over a community comes with the responsibility of not only maintaining the product but also engaging with the community and developing a strong relationship with it. In return, maintainers can be incentivized both through financial and non-financial incentives (e.g., finding friends, feedback, and a long-term future) to engage in this work.

In thinking about supporting OSS maintenance as a career, there are some critical research questions to consider before good answers can be provided. For example:
- How can leadership of a company be convinced to see the value of OSS software?
- How can OSS be sustained financially over longer periods of time?
- How can the workload of maintainers be reduced?
- How can social collaboration, such as work patterns and choices of technologies, be studied in order to address this question?
- How can research produce implementable results?

Answers to these questions can help provide a better understanding of the work and funding mechanisms for maintenance work in open source.

**OSS supply chains: How to support OSS supply chains and make them more visible?**

The support and visibility of OSS supply chains is multifaceted. First, it is **important to differentiate visibility from transparency** and acknowledge the importance of the two in ensuring a robust supply chain. While transparency of a supply chain is evident in the ease with which information on each supply chain node can be retrieved, visibility into a supply chain is evident in identifiability of metrics important to the users associated with a supply chain. Ensuring both is a critical aspect of recognizing how supply chains operate. Second, **recognizing the strength of sociotechnical relationships** embedded in a supply chain is an important consideration as well (e.g., the number of people involved in a supply chain and the kinds of relationships they share in terms of their dependency on each other and availability of options). Third, **acknowledging the legal responsibility of corporations and making it visible** has the potential to influence not only the visibility of supply chains but also the open source ecosystems as corporations signal trust or distrust through the decisions they make with regards to a community, a product, or other nodes in a chain. Finally, OSS supply chain visibility can help **manage the growth of OSS projects**. Per Lehman's laws of software evolution, software systems need to grow in order to sustain user satisfaction. However, with growth comes complexity, which challenges further growth. With the support of supply chains, however, **managing the growth of individual projects by outsourcing that growth to other nodes** in the chain becomes a relevant possibility (e.g., using dependencies instead of developing code in-house). But the extent to which this strategy could help is debatable as dependency management has its own challenges. Moreover, the knowledge of each node of a supply chain along with metrics important to its users can also be used by bad actors to disrupt supply chains. That is, there are important tradeoffs to consider.

## Panel #2

**Theme: How and to what extent can we rally stakeholders to consider broader concerns in open source ecosystems and which stakeholders are able to respond to and fulfill such responsibilities?**

This panel followed similar format as that of panel #1. That is, it began with a set of prompts for each panelist. After each panelist had addressed a prompt, the rest of the workshop participants had the opportunity to respond to that prompt or to the responses from the panelist. The list of prompts included:

- Can the actors in open source ecosystems be convinced and empowered to respond to critical collective concerns? Which actors? Why or why not?
- What are the roadblocks in the day-to-day work of volunteers, employees, users, and organizational actors that challenge individual or collective action?
- What does the mitigation of these roadblocks look like for actors (e.g., maintainers, organizations etc). In terms of actions they can take (e.g., policies, resources, etc.)?
- Where do policies fail in triggering meaningful action? Why?

Discussion around the prompts mentioned above led to the following discussion items around sustainability of open source software communities, the use of autonomous AI agents, and the need for maintenance-related work in open source:

From the perspective of a corporation intending to rally the stakeholders of a community towards shared goals, **sustainability of a community should be important.** Identifying that communities are often groups of people with a shared vision is a crucial aspect of community sustainability. More broadly, though, the question of rallying stakeholders is tied up with identifying the meaning of accountability. Determining collective goals, taking relevant actions and therefore having the ability to gauge their implications correctly is an accountability concern and is challenging to implement (e.g., due to the unpredictability of supply chain dependencies). Determining these goals and aligning communities with those goals is somewhat possible, however, anticipating corresponding outcomes and/or ramifications is difficult. For example, the visibility of source code in an open source project might give an incorrect impression that implications of collective decisions are sufficiently understood. But it would be a mistake to assume so. Not all aspects of community decisions are reflected through a project's codebase and there is much to be learned from the social and organizational dynamics in a community. Additionally, the meaning and importance of collective concerns is unclear and debatable. As collective concerns, product development and maintenance might be relatively easier to rally for, than community engagement. It can be challenging to empower people to consider and respond to community-engagement concerns that they may not experience on a day-to-day basis. When community engagement concerns are not well recognized and planned for, it may not be noticeable immediately but it essentially translates to 'death by a thousand cuts'. That is, failure of a community to collectively acknowledge what is important for their sustainability can lead to serious and unwanted community outcomes.

Another way to think about rallying for collective concerns is to **imagine members/stakeholders of OSS communities as autonomous agents**: What are the sources of influence for such agents given that communities are decentralized? These sources of influence are emergent when a sufficient number of agents begin to move in a certain direction that creates pressures on others to follow along. For example, there are fewer hypertoxic communities now than in the past because a collective shunning of such a culture has occurred. In terms of product development, when most community members tend to prefer writing code over maintenance-related work, rallying to get the initial development work done becomes relatively easier than later parts of project contribution work which includes much maintenance and refactoring work. Volunteers often do not show up to support that kind of collective effort. That is, **considering enforceable regulation on these community agents** (especially when industries fail to self-regulate), getting a **better understanding of their day-to-day concerns**, and **developing a language to speak to those day-to-day concerns** are some ways to initiate collective action among these agents.

However, there is an understandable conflict of ideologies in open source ecosystems which is a challenge to the ability of OSS communities to rally collective action. Some AWS engineers have been known to be **hesitant in participating in OSS** activities because they do not necessarily know if they are allowed to take specific actions. This hesitancy was related to the perception that their lack of experience in OSS might be counted against them and their contributions might not be accepted which highlights the importance of creating welcoming communities. Additionally, some kinds of work such as refactoring, community building, documentation, and other seemingly peripheral tasks that are important for OSS communities are not considered as valuable by the management of corporations. Given this, **it must be acknowledged that OSS value does not necessarily manifest through quarterly profit but through longer-term engagement**. Members of the Rust community have reported that their most important work in the community has been unplanned which has led to the success of Rust in many ways. **It is to be noted that when a sufficient number of people contribute significantly but in unplanned ways, it indicates that collective action does not necessarily need collective direction.** Community members could be working on very different, self-selected, unplanned, but important tasks that contribute significantly to the success of the project. In terms of rallying, giving members space to experiment and find their way around the product and the community, therefore, becomes incredibly important. In terms of roadblocks to rallying collective action, burn out, interpersonal conflict, lack of project management work are additional significant barriers.

Thinking of collective concerns also implores a consideration of project age because when projects age the **need for maintenance related work**, **as opposed to new feature development work, tends to be proportionately higher**. In order to get maintenance to be recognized as a collective concern in a community, it is crucial that most, if not all, contributors be willing to take on both kinds of tasks. However, contributors often want to work on newer projects/develop new features which is likely to encourage them to move on from older projects. It can then become challenging to get the maintenance of older projects be recognized as a collective concern. While some maintainers have been known to continue to maintain their projects out of a sense of obligation, they express frustration at the status quo and acknowledge wanting to move on. An important piece of recognizing this work as a collective concern and, therefore, creating a maintenance work force is to support a steady supply of maintenance-related skills and to train individuals to understand the relevance of tasks such as documentation, better curation of code, responding to user questions, posting community updates etc. Identifying such diverse skill sets and including non-developers as important candidates for maintenance positions can (a) help create such a workforce and avoid maintainer burn out; and (b) help young people and students get involved with open source communities. Contrary to the OSS projects in the past when there were fewer projects competing for talent, OSS projects these days are numerous and of higher complexity which makes it challenging for students and beginners to get involved. It is also important to demonstrate the business value of open source to younger generations who might be more interested in creating their start-ups. Some panelists expressed the belief that there is a lack of awareness among younger generations about the non-financial but significant career boosting incentives that OSS can provide (e.g., mentorship, impact on society, learning important skills, other long-term benefits)

To some extent, these **challenges related to project and community maintenance could be addressed with the help of AI agents.** For example, AI agents that are friendly chatbots and can respond quickly to poorly created PRs or can do code refactorings could help mitigate resource-related issues. This possibility begs questions about **the parts of a maintainer's work that AI can assist with** and about **the responsibility and accountability that could/would be assigned with AI** involved in maintenance work. What would future course of action for a group of maintainers look like when AI rejects PRs from senior maintainers? It is also important to acknowledge that **human maintainers might be incapable of responding to a barrage of AI generated PRs**. Thinking about **AI integration with maintainers' work flow as an addition or an upgrade to existing automation-related infrastructure** such as CI Dependabot may help provide some clarity around these issues given that Dependabot and AI are essentially doing similar kinds of tasks.

Another aspect of collective action is considering how **OSS communities can support individuals' careers and personal lives** in addition to considering how individuals can contribute to OSS. OSS projects are accountable to their contributors in making sure that they are satisfied and happy. OSS communities could provide mentorship and long-term relationships to individual contributors. It is also critical to acknowledge that OSS members often work with different communities at the same time and so the boundaries of membership, and therefore the expectations from it, are not necessarily as clear. Good OSS projects that invest time in documentation, provide good governance, establish communication channels such as Slack, create a sense of community and shared purpose etc. can serve as a model of baseline expectations that contributors could ask of their communities. Overall, it is important to sincerely and intentionally consider (a) how someone's first experience in an open source community might be and how that experience can be consistently improved over time; (b) that strong and sustainable communities are built not around projects but people, relationships, important problems that are meaningful and persist over time ; and (c) that we need a foundational vocabulary to articulate macro-level collective concerns of OSS ecosystems in a manner that makes sense in the context of the day-to-day work of community members (e.g., translating supply chain related concerns into more concrete questions relevant to open source maintainers fixing bugs and reviewing PRs on a daily basis).

## Brainstorming Session #2

The panel #2 discussions generated the following topics for breakout discussion sessions among the participants. Please refer to the Appendix for further details on the questions discussed during the brainstorming session.

**AI agents for open source: Which AI agents can be built to support open source software communities?**

Currently, AI agents are being used in AWS to make it easier for its employees to contribute to OSS. It has been made a part of employee workflow such as through ticketing and training process, and through suggesting possible PR-related decisions to human reviewers, etc. Ensuring that AI agents do not hallucinate or misguide their users, there are at least three types of project management agents that can potentially be useful for the OSS contributors: PR agents, CI/CD agents, and sunsetting agents. PR agents can help PR creators determine which reviewers, mentors, documentation etc. they need to consult, promote recent PRs on social media to attract more contributors, etc. CI/CD agents could integrate with CI/CD pipelines and with the workflow of developers in a manner that would make recommendations for developers, for example, about abandoned or archived dependencies, and possibly also fix such problematic dependencies. Sunsetting agents could help plan departure of contributors, transfer knowledge, develop a migration guide, document work, and possibly also inform users of these changes. Other possibilities include using agents to reduce contributor burn out by reminding them to take breaks and to structure complex conversation threads and by helping conduct meaningful user research and surveys.

**Paths to OSS leadership: What can open source projects do for maintainers' careers and how can they help junior contributors discover meaningful paths to leadership?**

**Identifying the value of open source and demonstrating it to early to mid-career contributors** is the first crucial step in enabling leadership pathways for new contributors in open source communities.

For example, **open source communities enable consistent social interaction and a trusted community of coworkers** in comparison to corporations. Employers these days rarely commit to their employees in the ways that they used to in the past. There are more incentives for employees to keep switching jobs due to continuously changing employer policies around work from home, salary increase, etc. Moreover, when employees have to take breaks from work for child care, the corporate system is structured in such a manner that it could lead to employees dropping out completely. As a result, employees are unable to find a community of professional colleagues they can rely on. Open source communities mitigate these issues of constrained social interaction and limited personal and professional growth. Such community-building facets of professional growth are important and valuable for a long-term engineering career. When employers make it challenging for their employees to continue to contribute to open source, employees have better incentives in leaving their current jobs and find ways to continue their association with existing coworkers in open source communities.

**Open source projects make commitment to newcomers through mentorship, training employable skillsets, etc.** Open source engagement helps build networks and find work. Early career engineers can use open source as a platform to develop their skills and transition to a paid role eventually. Participation in open source can help engineers build non-technical project management skills. However, open source mentors need to find new ways of supporting their mentees. For example, some options include demonstrating the connection between mentees' skills and the needs of employers, helping mentees understand the meaning of open source jobs in practice, providing a viable way to shift careers, providing career coaching and helping mentees think broadly about career progression and skill development, improving their visibility and skills to prospective employers etc. It is also important to acknowledge the cultural differences across open source communities and the ways in which OSS communities in other cultures support their contributors. Contributors in collectivist cultures are unable to raise their hands by themselves, be visible, and become effective contributors. The research and practice of open source has been, at least for the past 30 years, steeped in Western-oriented viewpoints on how open source works. Considering the cultural differences can help us learn more about the needs of mentees from various backgrounds and the ways in which they can be supported.

**Welcoming environment in OSS communities: What are the five actionable ways that a welcoming environment can be created in open source communities?**

- Intentional and proactive mentorship through providing office hours with no specific agenda and allowing free form discussion that enables people to ask questions and lower the barrier to participation.
- Actively recruiting people for open source projects who are particularly interested in and skilled at community engagement by developing personal relationships in open source communities, and promoting the non-technical benefits of engaging in open source.
- Actively recognizing contributors and encouraging them to focus on getting their second contribution in and following up on their first set of contributions in a manner that helps them stay engaged and feel supported instead of feeling overwhelmed and lost.
- Providing meaningful titles and roles to new contributors who ask for ways to get engaged and contribute to the community. Even though this might disincentivize others from stepping up in those roles when a certain responsibility is associated with a title ahead of time, but, it also can be a useful signal to a potential employer when contributors are able to label their contributions to open source with a meaningful title that employers can understand, relate to, and value. However, it is important to make sure that these titles are meaningful and difficult to earn.
- Rely on the possibility of using AI agents to help orient newcomers to community culture.

## Panel #3

**Theme: How can the stakeholders of open source ecosystems engage with each other? What are the ways in which meaningful interaction can be enabled with communities? And, what is the role of Open Source Program Offices (OSPOs) in doing the same?**

The panel began with a set of prompts for each panelist. In this panel, the workshop participants had the option to submit written questions via Slack during the panel and to ask follow up questions after the panel. The list of prompts for this panel included:

- What is your experience with OSPOs? How did you get involved with OSPOs - basically, what was your entry point?
- Could you discuss the role and inner workings of OSPOS i.e., institutionalizing collaboration and building interfaces between stakeholders.
- What are the goals of the collaborations enabled by OSPOs in academic and industrial settings? How do OSPOs prioritize them?
- What is the composition of the organizations, individuals, and other kinds of entities OSPOs connect? How do OSPOs make sense of their motivations?
- What are the challenges that OSPOs run into? What resources do they need?
- In an ideal world, what should OSPOs look like? Why?
- What would an exciting research agenda look like for you, around the topic of OSPOs and, more broadly, engagement across stakeholders of open source ecosystems?

Discussion around the prompts mentioned above led to the following discussion items around the need for creating OSPOs:

At Amazon, the Open Source Program Office (OSPO) was motivated by the legal liabilities of working with open source software, however, the OSPO was not established as a part of it. Currently, the work of OSPOs is not related to the business-line lawyers but only to the IP/Patent lawyers. To avoid liability and other costs, OSPOs work closely with most teams on the software development lifecycle to screen software dependencies and to ensure that corresponding communities that corporations are engaging with are viable, healthy, and sustainable. Regardless of the extent of work that OSPOs invest into making relevant connections among actors of open source software ecosystems, it remains a challenge to (a) characterize the work OSPOs do; and (b) demonstrate the business value of corporate OSPOs. One possibility is to define **OSPO work as that of engaging in diplomacy and trying to advocate for open source project support inside a company but also advancing company interests in open source communities.** For the purposes of academic research, OSPOs' role can also be understood as connecting scholars with developers working in corporate settings.

An important aspect of operating an Open Source Program Office is to **understand the difference between open source and free/libre software**. The work of OSPOs in corporations, therefore, demands careful navigation of and working with General Purpose Licensing software. Historically, Technology Transfer Offices (TTO) have considered open source as a bad strategic move as they view open source as giving away Intellectual Property. Therefore, it might seem that their involvement in OSPOs might cause some conflict and tension. However, current and ongoing experience of university OSPOs with TTOs, at least in the U.S., does not indicate the possibility of conflict.

## Brainstorming Session #3

The panel #3 discussions generated the following topics for breakout sessions among the participants. Please refer to the Appendix for further details on the questions discussed during the brainstorming session.

**Helping OSPOs: What can open source organizations (e.g., Apache) do to help OSPOs?**

When corporations need to use an open source project, their corresponding corporate OSPOs might need to find and work with commercial open source vendors. These vendors offer support, such as formal contracts, that ensure a corporation's needs (e.g., in relation to an open source project) are met within a specific time frame. **Open Source foundations can help maintain a list of vendors and service providers in ways that does not endorse them but keeps a track of their availability on a regular basis.** Such a list could help corporations choose open source projects based on the extent of vendor support available for them. Furthermore, vendors could assess project health reasonably well if projects are able to align with other projects on a set of metrics that can demonstrate its sustainability. While there are challenges to publicly publishing these metrics as they might hurt vendors, some foundations publish them publicly but deliberately avoid attracting attention to them. Amazon OSPO publishes these metrics internally because it can contain information and content not well suited for public distribution. If open source organizations such Apache can provide these metrics to OSPOs, that would enable OSPOs to compare different players in an open source ecosystem. However, it must be taken into consideration that different OSPOs measure value differently. Another way for open source organizations to support OSPOs is to fund mentors who might be serving as liaisons between OSPOs and open source projects. These liaisons could communicate the needs of a project further enabling OSPOs to make the right connections.

**Managing scaling of open source communities: How do norms change when projects scale and how can anthropologists help plan the transition?**

The growth of open source projects is challenging to navigate. **As a village becomes a town and a town becomes a city, the internal culture of implicit values and norms gives way to explicit rules and regulations, and transferring leadership becomes challenging.** This is one of the reasons why some open source contributors prefer to work with smaller communities that seem less bureaucratic. Per the German sociologist and philosopher Georg Simmel, a fundamental shift occurs within the social dynamics and complexity of a group of people when it grows from 2 to 3 members, since a member of the original group could then be overruled. Furthermore, when communities grow from 50 –still a manageable group of people that might respect their own norms– to 100 members, norms tend to get replaced by rules. Norms may no longer be obvious, clear or even respected. In particular, transferring knowledge among community members might vary across different project stages: As the first generation of maintainers pass on their knowledge to the second generation, the third generation could lose access to parts of that knowledge which the second generation of maintainers does not care about. Additionally, early maintainers and members that are not comfortable with bigger communities might consider leaving a community as it grows which leads to knowledge loss. An important way to navigate such issues related to scaling is to understand and learn how other communities and projects have handled the same. That is, doing case studies, sharing those case studies with students and early contributors, and finding ways to learn both deeply about a few case studies and also broadly across a number of case studies.

## Takeaways: Important and Urgent Questions

The participation of both industry professionals and academics in the workshop resulted in exciting new questions, reiterated some of the known issues and pain points, and highlighted how certain research questions have not translated well to practice settings. It was acknowledged by the participants that this workshop was a helpful joint forum to recognize common, important, and urgent questions for both practitioners and academics. We discuss these questions below. For key themes from the workshop, please refer to the executive summary.

**Maintainer Support**
- It is becoming increasingly crucial to recognize the kinds of funding mechanisms that can support different kinds of software development, for example, feature development versus long-term software maintenance. What are the funding mechanisms that can support long-term maintenance work whose impact is often diffused and therefore difficult to measure?
- How to engage maintainers in staying or getting involved in an OSS project?
- How to define their tasks and responsibilities? What are maintainers accountable for and what should be expected of them?
- How to plan for smooth maintainer transitions and frictionless scaling of OSS communities?
- What are the best practices for a developer/contributors retiring from an OSS project and how can technologies help?
- How can service to OSS be integrated as a part of job description in corporations similar to how academic service is a part of an academic's job?

**AI And Open Source**
- What would the role of maintainers and community members look like when future AI agents take up the bulk of work? How would it shape the maintenance and collaboration culture of a community?
- What kinds of organizational processes might be relevant for communities working on open source AI models?
- What processes of OSS apply or don't apply to communities around open AI models?

**Scaling OSS**
- Why do some communities scale gradually and others don't?
- What kinds of structures/norms/processes do communities need to build as they grow?
- What specific community practices are correlated with community growth and sustainability?
- How can we characterize the needs of a community as it grows and recognize possible friction points? For example, how can we identify a village that hasn't realized that it has become a town?

**Open Source Value And Risks**
- How can we determine the monetary and non-monetary value of OSS in different contexts such as OSS business strategy and/or external contributions?
- How do corporations generate value from open source?
- What is the correlation between corporate engagement in FLOSS and profitability?
- What disclosures about OSS are necessary to make it accountable?
- How can we identify identities/actors in OSS for the purposes of accountability? (e.g., people, orgs, AI)
- How can we think and record the value of OSS work practices that cannot be measured by traditional means? Technical work that cannot be measured is not valued similarly as other measurable tasks that have clear metrics associated with it. For example, measuring the effort that maintainers invest in long-term maintenance of OSS projects, or the work of open source diplomats that operate based on their relationship with multiple individuals and corporations.

**OSS Metrics**
- Which metrics predict community failures like forks or abandonments?
- Metrics can be problematic. How can metrics be generated and used in a responsible manner?

**OSS Engagement**
- How can we continue the conversation on accountability in OSS and also bring in more voices?
- How can a global perspective on OSS communities be developed that encompasses both Western and non-Western perspectives? How can people around the world be brought on board on largely Western open source projects?
- What differentiates successful OSPOs from unsuccessful OSPOs?
- How can academic research be informed by industry experience so that research agendas can benefit both the practitioners and the researchers?
- How can academics get involved in open source practice, in addition to open source research, in order to help the practitioners better understand OSS ecosystems? Such involvement of academics in practitioner-oriented settings is increasingly needed.

As with most generative conversations, this workshop helped highlight important questions about the current and future needs of open source communities. These questions have surfaced as precursors to thinking and articulating relevant accountability issues which are complex and diverse. Our hope is that these questions can catalyze new pathways for both researchers and practitioners to explore how accountability could manifest in various contexts. We are sincerely grateful to the workshop participants for joining us for the workshop and sharing their thoughts and expertise.

## Acknowledgements

This workshop was funded by the National Science Foundation (NSF) under Grant No. 2317168 and 2317169.

## Cite

Sharma, N., Bock, T., Bowen, R., Choudhury, S., Fitzgerald, B., Germonprez, M., Herbsleb, J., Howison, J., Hughes, T., Lee, M. K., Lieggi, S., Liesenfeld, A., Link, G., Matsakis, N., Mockus, A., Ramasubbu, N., Robinson, C., Robles, G., Ruff, N., … Yoo, C. (2026). *Accountability in Open Source Software Ecosystems: Workshop Report*. Workshop on Accountability and Open Source Software, Pittsburgh, PA, USA. Carnegie Mellon University; University of Pittsburgh.
https://www.cmu.edu/s3d/open-source-communities-and-accountability-workshop/workshop-report-final.pdf

# Appendix

## Panel programming

**Panel Discussion #1: Who are the stakeholders of open source software communities? How can we identify their needs?**

**Moderator:** [Jim Herbsleb](), Professor, Software and Societal Systems Department, School of Computer Science, Carnegie Mellon University

**Panelists:**

- [Brian Fitzgerald](): Krehbiel Chair in Innovation in Business & Technology, University of Limerick
- [Matt Germonprez](): Mutual of Omaha Distinguished Chair of Information Science & Technology, University of Nebraska; Co-founder of the Linux Foundation Community Health Analytics OSS Project (CHAOSS)
- [Nicholas Matsakis](): One of the long-time leaders of the Rust project; Co-lead Rust Language Team; Senior Principal Engineer at Amazon
- [Christopher Yoo](): Imasogie Professor in Law and Technology, University of Pennsylvania

**Panel Discussion #2: How can we rally stakeholders, including OSS communities, to consider broader concerns such as OSS supply chain security, performance, and API feature requirements etc. To what extent do we need to do that? Which stakeholders are able to respond to and fulfill such responsibilities?**

**Moderator:** [Sonali Shah](): Professor & May Faculty Fellow, Gies College of Business, University of Illinois at Urbana-Champaign
**Panelists**:
- [Rich Bowen](): Open Source strategist at AWS
- [Audris Mockus](): Ericsson-Harlan Mills Chair Professor, Department of Electrical Engineering and Computer Science, University of Tennessee
- [Nandini Sharma](): Postdoctoral Associate, Carnegie Mellon University

**Panel Discussion #3: How can the stakeholders of open source ecosystems engage with each other? What are the ways in which meaningful interaction can be enabled with communities? What is the role of OSPOs in doing the same?**

**Moderator:** [Georg Link](): Co-founder of the Linux Foundation Community Health Analytics OSS (CHAOSS) Project
**Panelists**:
- [Sayeed Choudhury](): Director of the Open Source Programs Office (OSPO) at Carnegie Mellon Libraries.
- [Stephanie Lieggi](): Executive Director, CROSS (Center for Research in Open Source Software), UC Santa Cruz
- [Nithya Ruff](): Head of the Amazon Open Source Program Office
- [Stephen Walli](), Principal Program Manager at the Azure Office of the CTO, Microsoft, Open Source Advocate

# Brainstorming Discussions

**Brainstorming Session #1**
**Questions:**
- **Group#1:** Accountability as a baseline standard for selecting a component into a commercial product; avoiding liability.
- **Group#2:** Open source maintenance as a career: How to make it happen? Sustainable funding; What is a reasonable expectation of a maintainer of an OSS project?
- **Group#3:** How to make multiple front doors for organized groups to engage with open source?
- **Group#4:** How can we make OSS supply chains more visible?

**Brainstorming Session #2**
**Questions:**
- **Group#1:** What are the top 10 AI agents we must build to help open source development?
- **Group#2:** What can open source projects do for contributors' careers? What is their path to leadership? How do we think about the role of open source in people's careers? What do contributor pathways look like for mid-to-late stage contributors? What is their path to leadership?
- **Group#3:** What is the best way (5 actionable items) for open source projects to have a welcoming environment? How do we expand the scope of participants beyond software developers and non-code contributors?

**Brainstorming Session #3**
**Questions:**
- **Group#1:** What can open source organizations (e.g., Apache) do to help your role as an OSPO?
- **Group#2:** What are the conflicts between OSPOs and the business units and how are they managed? How do you get lawyers to help you? Beyond supporting OSS, what is the role of OSPOs in adjusting company strategy to be more open?
- **Group#3:** What are the ways in which organizational imaginaries can be built and expanded beyond existing understandings?
- **Group #4:** What happens when a "village becomes a town which becomes a city"? How do norms and rules change and shape the sociology of a tech company/open source communities? How can anthropologists help/plan this transition?

## Participants

| Name | Affiliation |
| --- | --- |
| Andreas Liesenfeld | Radboud University Nijmegen, The Netherlands |
| Audris Mockus | University of Tennessee, Knoxville, USA |
| Bogdan Vasilescu | Carnegie Mellon University, USA |
| Brian Fitzgerald | Lero and University of Limerick, Ireland |
| Christopher Robinson | OpenSSF/Linux Foundation, USA |
| Christopher Yoo | University of Pennsylvania, USA |
| Georg Link | Bitergia, Spain |
| Gregorio Robles | Universidad Rey Juan Carlos, Spain |
| Igor Steinmacher | Northern Arizona University, USA |
| James Howison | The University of Texas at Austin, USA |
| Jim Herbsleb | Carnegie Mellon University, USA |
| Matt Germonprez | University of Nebraska Omaha, USA |
| Min Kyung Lee | The University of Texas at Austin, USA |
| Nandini Sharma | Carnegie Mellon University, USA |
| Narayan Ramasubbu | University of Pittsburgh, USA |
| Nicholas Matsakis | Amazon, Rust, USA |
| Nithya Ruff | Amazon and Linux Foundation, USA |
| Rich Bowen | Amazon Web Services, USA |
| Sayeed Choudhury | Carnegie Mellon University, USA |
| Sonali Shah | University of Illinois at Urbana-Champaign, USA |
| Stephanie Lieggi | University of California Santa Cruz, USA |
| Stephen Walli | Microsoft, USA |
| Thomas Bock | Carnegie Mellon University, USA |
| Tom Hughes | Carnegie Mellon University, USA |